\documentclass[conference]{IEEEtran}
\IEEEoverridecommandlockouts
\usepackage{cite}
\usepackage{amsmath,amssymb,amsfonts}
\usepackage{algorithmic}
\usepackage{graphicx}
\usepackage{textcomp}
\usepackage{xcolor}
\usepackage{subcaption}
\usepackage{booktabs}
\usepackage{placeins}
\usepackage{float}
\usepackage{stfloats}

\def\BibTeX{{\rm B\kern-.05em{\sc i\kern-.025em b}\kern-.08em
    T\kern-.1667em\lower.7ex\hbox{E}\kern-.125emX}}

\begin{document}

\title{Noise as a Double-Edged Sword: Reinforcement Learning Exploits Randomized Defenses in Neural Networks}

\author{
\IEEEauthorblockN{Steve Bakos}
\IEEEauthorblockA{\textit{Faculty of Science} \\
\textit{Ontario Tech University} \\
Oshawa, Ontario, Canada \\
steven.bakos@ontariotechu.net}
\and
\IEEEauthorblockN{Pooria Madani}
\IEEEauthorblockA{\textit{Faculty of Science} \\
\textit{Ontario Tech University} \\
Oshawa, Ontario, Canada \\
pooria.madani@ontariotechu.ca}
\and
\IEEEauthorblockN{Heidar Davoudi}
\IEEEauthorblockA{\textit{Faculty of Science} \\
\textit{Ontario Tech University} \\
Oshawa, Ontario, Canada \\
heidar.davoudi@ontariotechu.ca}
}
\maketitle

\begin{abstract}
This study investigates a counterintuitive phenomenon in adversarial machine learning: the potential for noise-based defenses to inadvertently aid evasion attacks in certain scenarios. While randomness is often employed as a defensive strategy against adversarial examples, our research reveals that this approach can sometimes backfire, particularly when facing adaptive attackers using reinforcement learning (RL). Our findings show that in specific cases, especially with visually noisy classes, the introduction of noise in the classifier's confidence values can be exploited by the RL attacker, leading to a significant increase in evasion success rates. In some instances, the noise-based defense scenario outperformed other strategies by up to 20\% on a subset of classes. However, this effect was not consistent across all classifiers tested, highlighting the complexity of the interaction between noise-based defenses and different models. These results suggest that in some cases, noise-based defenses can inadvertently create an adversarial training loop beneficial to the RL attacker. Our study emphasizes the need for a more nuanced approach to defensive strategies in adversarial machine learning, particularly in safety-critical applications. It challenges the assumption that randomness universally enhances defense against evasion attacks and highlights the importance of considering adaptive, RL-based attackers when designing robust defense mechanisms.
\end{abstract}


\section{Introduction}
Adversarial Machine Learning (AML) is a field focused on exposing and exploiting vulnerabilities in Machine Learning (ML) systems. From evading detection to compromising models, AML covers a wide range of attacks, each targeting different stages and parts of the ML pipeline. As ML systems become more integrated into critical applications such as autonomous vehicles, healthcare, and cybersecurity, their security becomes paramount \cite{lin2021adversarial}.

\begin{figure}[b]
\centering
\begin{subfigure}[b]{0.3\linewidth}
    \centering
    \includegraphics[width=\linewidth]{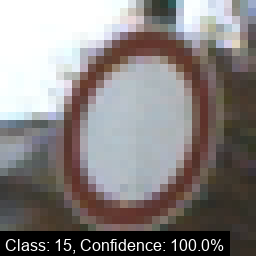}
    \caption{}
    \label{fig:classified_image}
\end{subfigure}
\hfill
\begin{subfigure}[b]{0.3\linewidth}
    \centering
    \includegraphics[width=\linewidth]{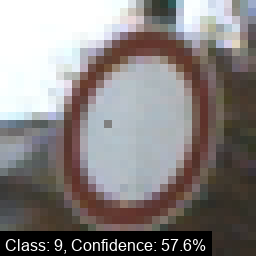}
    \caption{}
    \label{fig:misclassified_image}
\end{subfigure}
\hfill
\begin{subfigure}[b]{0.3\linewidth}
    \centering
    \includegraphics[width=\linewidth]{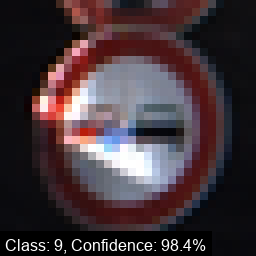}
    \caption{}
    \label{fig:true_class_image}
\end{subfigure}
\caption{(a) Original image correctly classified with 100\% confidence. (b) Modified image misclassified with 57.6\% confidence after a single pixel modification. (c) True Class 9 image.}
\label{fig:classification_example}
\end{figure}

\begin{figure}[htbp]
\centering
\includegraphics[width=0.8\linewidth]{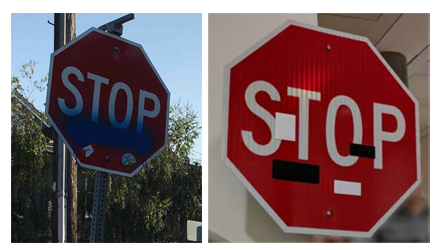}
\caption{Example of a physical-world adversarial attack on stop signs. Left: A stop sign with actual graffiti. Right: A stop sign with carefully designed stickers applied to mimic graffiti, capable of fooling image classifiers. These stickers represent a real-world adversarial attack, not digital manipulation. (Image source: \cite{eykholt2018robust})}
\label{fig:stop_sign_attack}
\end{figure}

Evasion attacks, where adversaries attempt to fool trained models during inference by subtly modifying input data, have been a central focus in AML research. These attacks pose significant challenges, especially in safety-critical domains where misclassification can have severe consequences. For example, Fig. \ref{fig:classification_example} illustrates the vulnerability of ML models to subtle evasion attacks, where minor modifications can lead to misclassification. As depicted in Fig. \ref{fig:stop_sign_attack}, evasion attacks can extend to the physical world, posing significant challenges for applications like autonomous vehicles. 

To counter evasion attacks, various defensive strategies have been proposed. One popular approach is the use of noise-based defenses, which introduce randomness into the classification process. The intuition behind this strategy is that randomness can make it harder for attackers to reliably generate adversarial examples. For instance, \cite{miller2014adversarial} demonstrated that randomness in data selection could defend against targeted attacks while maintaining performance in non-adversarial settings.

However, the effectiveness of noise-based defenses against adaptive attacks, particularly those employing reinforcement learning (RL), remains an open question. Recent work has begun to explore the application of RL techniques to create more sophisticated and adaptive adversarial attacks \cite{10208834, yang2020patchattack}. These RL-based approaches have shown promise in overcoming some limitations of static attack methods, presenting new challenges for existing defense strategies. Our research extends this concept by exploring how RL-based attackers can exploit noise-based defenses designed to prevent such attacks.

Our work aims to bridge this gap by systematically evaluating the impact of noise-based defenses against an RL-based adversarial agent. Specifically, we investigate how varying levels of information disclosure and the introduction of noise in a classifier's confidence values affect an RL agent's ability to develop successful evasion attacks. This work's contributions are threefold:
\begin{enumerate}
    \item We conduct a systematic evaluation of how varying levels of information disclosure and noise-based defenses affect an RL agent's ability to develop successful evasion attacks across multiple deep-learning classifier architectures.
    
    \item We demonstrate the counter-intuitive impact of noise-based defenses, revealing scenarios where introducing noise in the classifier's confidence values can aid the RL-based adversarial agent for certain classifiers and image classes.
    
    \item We provide a comprehensive benchmark for assessing the effectiveness of noise-based defensive strategies against adaptive RL attackers in AML, highlighting the variability of outcomes across different classifier architectures.
\end{enumerate}

The remainder of this paper is organized as follows. Section II details the background and related work. Section III discusses the experimental setup and methodology used. Section IV presents and discusses the results of the experiments, including the performance of each defensive strategy. Section V provides a discussion of our findings. Section VI discusses the limitations of this work. Finally, Section VII combines future work and conclusions, summarizing the key findings, discussing their implications for noise-based defenses in AML, and outlining directions for future research.

\section{Related Work}
\subsection{Evasion Attacks}
Evasion attacks, where adversaries attempt to fool trained models during the inference phase, have been a central focus in adversarial machine learning research. These attacks aim to cause misclassification by making subtle modifications to input data. 

One of the seminal works in this area is the Fast Gradient Sign Method (FGSM) introduced by Goodfellow et al. \cite{goodfellow2015explaining}. FGSM generates adversarial examples by using the sign of the gradient of the loss with respect to the input, allowing for quick generation of adversarial examples. Building on this, Kurakin et al. \cite{kurakin2016adversarial} proposed an iterative version of FGSM for more sophisticated attacks.

Another significant advancement came with the Projected Gradient Descent (PGD) attack, introduced by Madry et al. \cite{madry2019deep}. PGD is considered one of the strongest first-order attacks and has become a standard benchmark for adversarial robustness. It operates by performing multiple steps of FGSM and projecting the perturbations onto a specified norm ball, resulting in more powerful adversarial examples. Madry et al. not only proposed PGD as an attack method but also demonstrated its effectiveness in adversarial training, showing that models trained against PGD attacks exhibit strong resistance to a wide range of other attacks.

The effectiveness of these attacks has been demonstrated across various domains. For instance, Eykholt et al. \cite{eykholt2018robust} showed that physical-world attacks on traffic signs are possible, highlighting the real-world implications of these vulnerabilities. Similarly, Su et al. \cite{su2019onepixel} demonstrated that even changing a single pixel in an image can be sufficient to fool state-of-the-art classifiers.

Recent work has begun to explore the application of reinforcement learning (RL) techniques to create more adaptive and sophisticated adversarial attacks. This approach allows for the development of attack strategies that can adapt to different models and defense mechanisms.

Yang et al. \cite{yang2020patchattack} demonstrated the use of RL in adversarial attacks by training an RL agent to add adversarial patches to images. Their approach, PatchAttack, showed that RL could be used to develop attack strategies that are both effective and less detectable than traditional methods.

Sarkar et al. \cite{10208834} proposed using RL agents to generate adversarial examples, specifically leveraging the confidence values output by the target classifier. Their method employed a sensitivity analysis using Gaussian noise masks across the image to determine which patches led to the greatest drop in classification confidence. This approach, while effective, inherently relies on access to the classifier's confidence values, both for the sensitivity analysis and in their reward function. Such a method would not be applicable in a true black-box scenario where only the predicted label is available.

These works highlight a growing trend in AML research: the use of RL to create more dynamic and adaptable attack strategies. By leveraging RL, attackers can potentially overcome some of the limitations of static attack methods, presenting new challenges for defensive strategies. However, the level of information required by these methods varies, with some approaches being more suitable for scenarios with greater access to model outputs than others.

Our work extends this line of research by exploring how an RL-based attacker can operate under varying levels of information, including a true black-box scenario where only the predicted label is available. This allows us to investigate the effectiveness of RL-based attacks across a broader range of real-world scenarios, including those with limited access to model internals.

\subsection{Noise-Based Defenses in Adversarial Machine Learning}

Noise-based defenses have emerged as a promising approach to protect ML models against adversarial attacks. The core idea is to introduce randomness into the model or its inputs, making it harder for attackers to craft effective adversarial examples. 

One notable work in this area is by Lecuyer et al. \cite{lecuyer2019certifiedrobustnessadversarialexamples}, who proposed a certified defense mechanism based on randomized smoothing. Their approach adds Gaussian noise to inputs during both training and inference, providing provable robustness guarantees against $l_2$-norm bounded perturbations.

Similarly, Liu et al. \cite{liu2018robustneuralnetworksrandom} introduced random transformations as a defense against adversarial examples. By applying random rescaling and padding to input images, they showed improved robustness against a range of attack methods.

However, the effectiveness of these noise-based defenses against adaptive attackers, particularly those using reinforcement learning, has not been thoroughly explored. Our work aims to fill this gap by systematically evaluating how an RL-based attacker can potentially exploit the randomness introduced by these defenses.

\section{Methodology}

\subsection{Background} 
Reinforcement Learning (RL) is a paradigm in machine learning where an agent learns to make decisions by interacting with an environment. Unlike supervised learning, RL does not rely on labeled datasets. Instead, the agent learns through trial and error, receiving rewards or penalties based on its actions \cite{sutton2018reinforcement}. In RL, the agent's goal is to maximize its cumulative reward over time by learning a policy that maps states to actions. Through iterative interactions with the environment, the agent continuously refines its policy to prioritize actions that yield higher cumulative rewards, thereby optimizing its decision-making strategy over time.

RL problems are typically formulated as a Markov Decision Process (MDP) \cite{garcia2013markov}, defined by a tuple $(S, A, P, R, \gamma)$, where $S$ is the set of possible states, $A$ is the set of possible actions, $P$ is the transition probability function, $R$ is the reward function, and $\gamma$ is the discount factor. At each time step $t$, the agent observes the current state $s_t \in S$, takes an action $a_t \in A$, receives a reward $r_t$, and transitions to a new state $s_{t+1}$ according to the probability $P(s_{t+1}|s_t, a_t)$.

The RL agent must adopt a policy optimization technique to discover sequences of actions that lead to higher rewards. Proximal Policy Optimization (PPO) \cite{schulman2017proximal} is a popular RL algorithm known for its stability and effectiveness. PPO belongs to the class of policy gradient methods, which directly optimize the policy. It improves upon earlier methods by using a clipped surrogate objective
\begin{equation}
L^{CLIP}(\theta) = \hat{\mathbb{E}}_t[\min(r_t(\theta)\hat{A}_t, \text{clip}(r_t(\theta), 1-\epsilon, 1+\epsilon)\hat{A}_t)],
\end{equation}
where $r_t(\theta)$ is the ratio of the probability of taking action $a_t$ under the new policy to that under the old policy, $\hat{A}_t$ is an estimator of the advantage function, and $\epsilon$ is a hyperparameter. This clipping mechanism prevents excessively large policy updates, leading to more stable training.

\subsection{Threat Model}
Let $f_\theta$ denote a pre-trained classification model with which an adversary can interact. The adversary can query $f_\theta$ with an input instance $x$ and receive $f_\theta(x)$. The adversary's goal is to find a perturbing noise $\delta$ such that:

\begin{equation}
\label{eq:adversarial_goal}
argmax f_\theta(x + \delta) \neq argmax f_\theta(x)
\end{equation}

We assume that the adversary has varying levels of access to the output of $f_\theta$, which we evaluate through four attack scenarios: (1) Black Box, where the classifier returns only the predicted label without confidence values; (2) True Distribution, where the classifier returns the predicted label along with correct confidence values for all classes; (3) True Confidence, Others Randomized, where the classifier returns the predicted label and correct confidence for that label while randomizing others; and (4) Correct Confidence Only, where the classifier returns the predicted label with its correct confidence value.

In the third scenario, which implements a noise-based defense, we employ a randomization technique to obfuscate the true confidence values of incorrect classes. This process preserves the confidence value for the correct class while distributing the remaining probability mass among other classes using a Dirichlet distribution. This approach maintains the model's accuracy for legitimate inputs while potentially confusing adversarial attempts to exploit confidence values of incorrect classes.

The choice of this randomization technique is justified by several factors. Firstly, it preserves the true confidence of the correct class, maintaining some level of fidelity to the original model output. Secondly, the use of the Dirichlet distribution ensures that the randomized probabilities still sum to one, maintaining a valid probability distribution. Thirdly, the uniform parameterization in the Dirichlet distribution avoids introducing any systematic bias towards particular incorrect classes. Finally, this approach creates a challenging environment for the adversary, as the confidence values for incorrect classes vary randomly between queries, even for the same input, while still providing consistent and accurate information for the correct class.

\begin{figure*}[b] 
\centering
\includegraphics[width=0.8\linewidth]{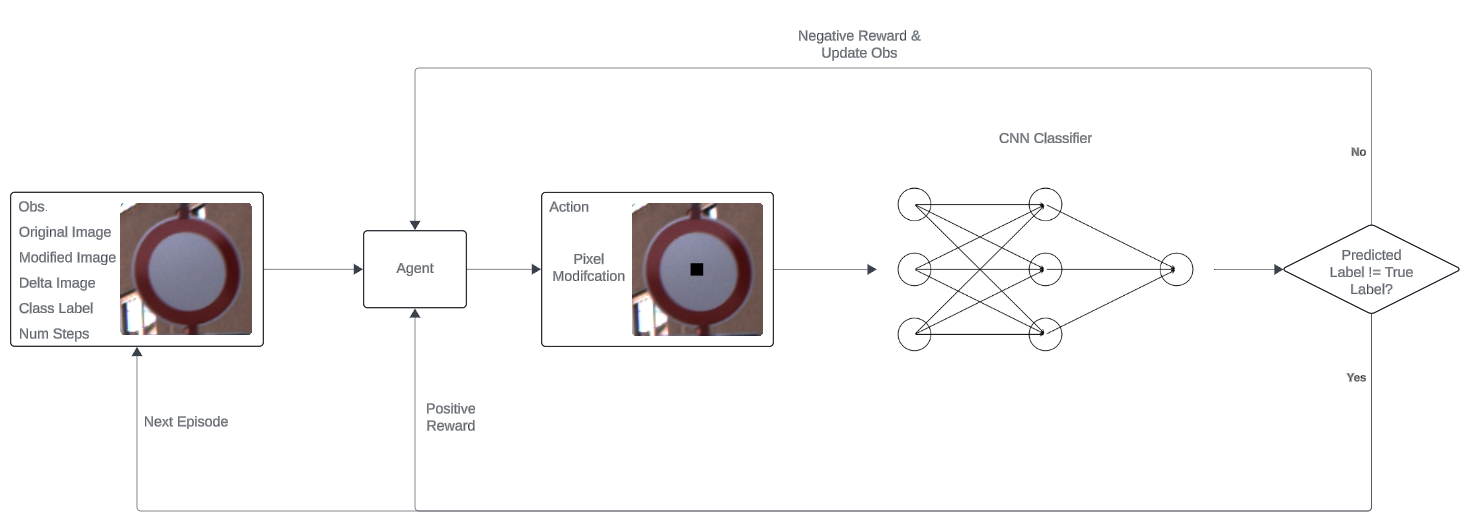}
\caption{Flow chart of the RL environment, showing the interaction between the agent, the environment, and the CNN classifier.}
\label{fig:environment_flowchart}
\end{figure*}

The adversary employs our proposed PPO-based RL attack methodology to find $\delta$ for a given $x$ and $f_\theta$, utilizing the different categories of information available in each scenario. We aim to measure the impact of the degree of informational feedback provided to the adversary by $f_\theta$ in terms of the discovered $\|\delta\|$ and the number of queries required to solve Equation \eqref{eq:adversarial_goal}.

\subsection{The RL-based Attack Agent}
The RL environment is designed to mimic the interaction between an attacker and the classifier. Fig. \ref{fig:environment_flowchart} illustrates the flow of information and actions within this environment.


Each episode begins with the selection of a random class from the dataset. An image from this class is chosen and presented to the classifier. If the classifier correctly predicts the label, the episode begins; otherwise, a new image is selected. The agent then receives an observation containing the original image, class label, the current modified image, the number of steps taken, and the delta image (difference between original and modified). Based on this observation, the agent selects an action, which is a pixel modification in the 32x32 image. The modified image is then passed to the CNN classifier. If the classifier's prediction matches the true label, the process continues with a negative reward, and the observation is updated. If misclassification occurs, the agent receives a positive reward, and a new episode begins. The agent can modify up to 32 pixels in the image (approximately 3\% of the total pixels) before the episode ends.

The agent's observation space depends on the attack scenario, including various levels of confidence information based on the scenario. The action space allows for the modification of a single pixel within the 32x32 image, with 8 possible intensity values, totaling 8,192 possible actions at any timestep.

The reward function is designed to encourage minimal modifications while successfully fooling the classifier. The agent receives -1 reward for each action taken, +10 reward if it successfully fools the classifier, and -10 reward if it fails to fool the classifier by the end of the episode. To ensure parity across all scenarios, the confidence values returned from the classifier are not used to influence the reward.

\section{Experimental Setup}

We use pre-trained ResNet18 \cite{he2015deepresiduallearningimage}, DenseNet121 \cite{huang2018denselyconnectedconvolutionalnetworks}, and MobileNetV2 \cite{sandler2018mobilenetv2} classifiers fine-tuned on the German Traffic Sign Recognition Benchmark (GTSRB) dataset \cite{Stallkamp2012} as $f_\theta$. The GTSRB dataset contains over 50,000 images of traffic signs across 43 classes. During preprocessing, images are resized to 32x32 pixels and normalized using the mean [0.485, 0.456, 0.406] and standard deviation [0.229, 0.224, 0.225] based on the ImageNet dataset.

ResNet18, containing about 11.2 million parameters, employs skip connections that enable the network to learn residual functions, addressing the vanishing gradient problem in deep networks. DenseNet121, with approximately 7 million parameters, is characterized by dense connectivity between layers, allowing for efficient parameter use and improved gradient flow. MobileNetV2, designed for efficiency with roughly 2.3 million parameters, uses depthwise separable convolutions and inverted residuals with linear bottlenecks, significantly reducing computational cost.

All classifiers achieve high accuracy on both training and test sets. ResNet18 achieves 99.61\% accuracy on the training set and 97.03\% on the test set. DenseNet121 shows 98.02\% accuracy on the training set and 94.49\% on the test set. MobileNetV2 attains 99.63\% accuracy on the training set and 97.01\% on the test set. These high accuracy rates provide a strong baseline for our adversarial attack experiments.

The PPO agent was implemented using PyTorch and Stable Baselines 3 \cite{stable-baselines3}, with the environment implemented using Gymnasium \cite{towers_gymnasium_2023}. Training was conducted on an Intel(R) Xeon(R) W-2223 CPU @ 3.6 GHz with 8 cores, 32 GB RAM, and an NVIDIA RTX A4000 GPU. To enhance training efficiency and robustness, we utilized 4 parallelized environments, each assigned a unique seed ranging from 42 to 45, ensuring diverse initial conditions and promoting better generalization of the learned attack strategies.

After experimenting with different hyperparameters, we found a set of parameters that worked well for the PPO agent across all classifier models. These include a learning rate of 3e-4, entropy coefficient of 0.01, clip range of 0.2, and other standard PPO parameters. All preprocessing and training of both the classifiers and agents were done using a fixed random seed of 42 to ensure reproducibility.

\subsection{The Training Process and Evaluation Metrics}

For each classifier model, the agent is trained under four different attack scenarios: Black Box, True Distribution, True Confidence with Others Randomized, and Correct Confidence Only. The training process for each scenario and model combination lasts approximately 24 hours, with the agent able to modify up to 32 pixels per image before the episode terminates.

We evaluate the performance of the adversarial agent using two key metrics:
\begin{itemize}
    \item \textbf{Lifetime Success Rate (LSR)}: This metric represents the rate at which the agent successfully fools the classifier over the entire course of its training. It is calculated as the number of successful episodes (where the agent successfully caused misclassification) divided by the total number of episodes attempted.
    
    \item \textbf{Average Actions to Fool (AAF)}: This metric measures the average number of pixel modifications required to induce misclassification within successful episodes. It provides insight into the efficiency of the attack strategy, with lower values indicating a more effective approach.
\end{itemize}

These metrics allow us to comprehensively assess the effectiveness of the RL-based attack strategy across different scenarios and identify potential vulnerabilities in the classifiers and noise-based defense mechanisms.

\section{Results}

\subsection{Overall Performance Across Attack Scenarios and Classifiers}

\begin{figure}[t]
\centering
\begin{subfigure}[b]{0.45\textwidth}
    \caption{ResNet18}
    \label{fig:resnet18_lifetime}
    \includegraphics[width=\textwidth]{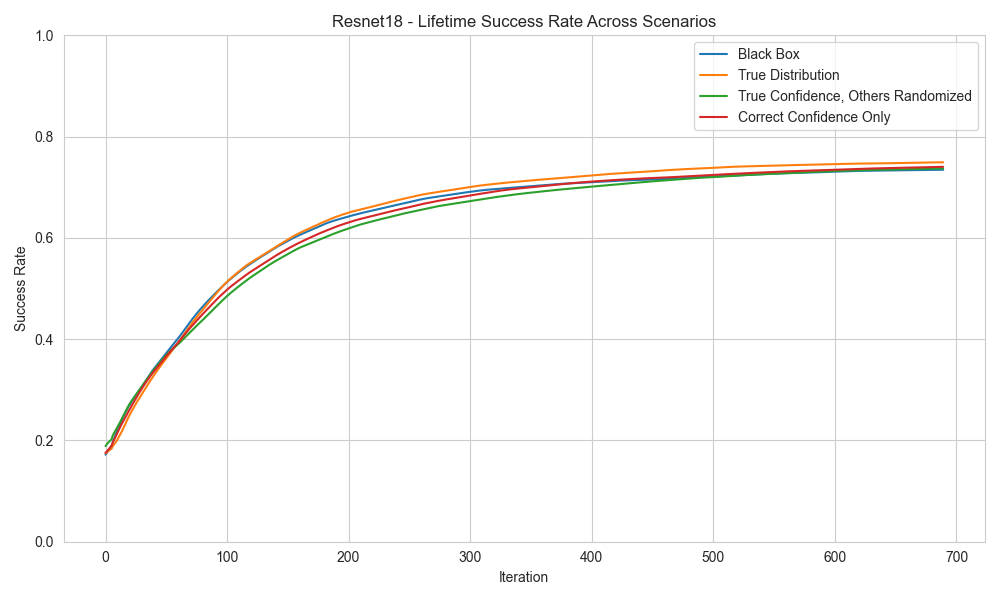}
\end{subfigure}

\begin{subfigure}[b]{0.45\textwidth}
    \caption{DenseNet121}
    \label{fig:densenet121_lifetime}
    \includegraphics[width=\textwidth]{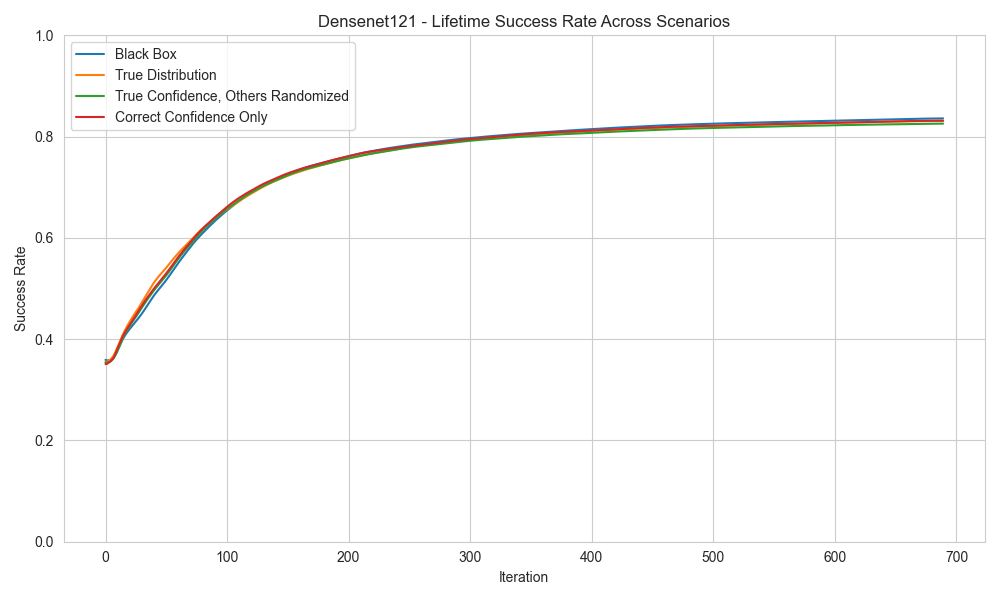}
\end{subfigure}

\begin{subfigure}[b]{0.45\textwidth}
    \caption{MobileNetV2}
    \label{fig:mobilenetv2_lifetime}
    \includegraphics[width=\textwidth]{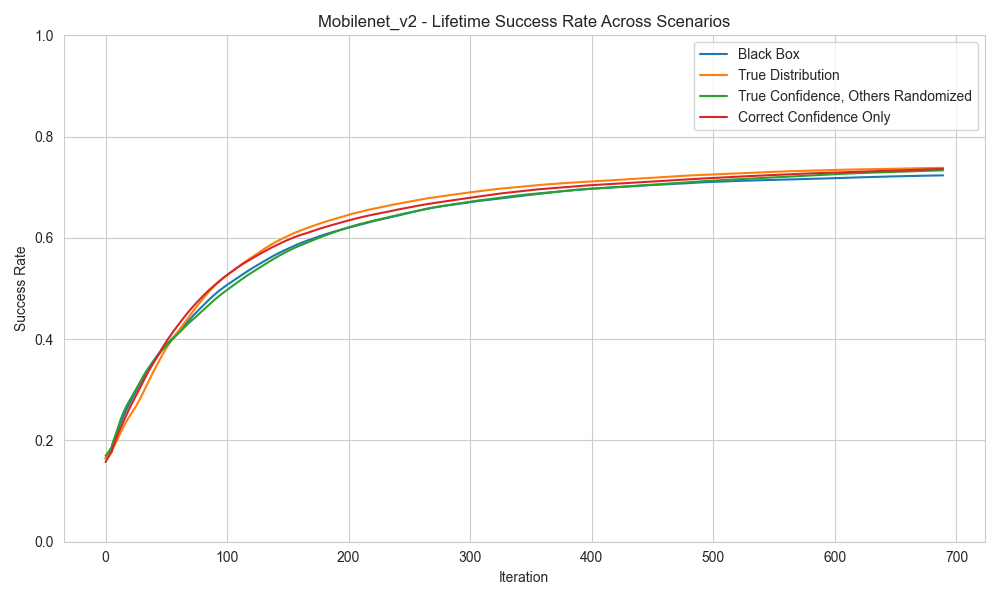}
\end{subfigure}
\caption{Averaged LSRs over the 43 classes for different attack scenarios across classifiers.}
\label{fig:lifetime_success_rates}
\end{figure}

Our analysis of the averaged LSRs across the 43 classes for each attack scenario, illustrated in Fig. \ref{fig:lifetime_success_rates}, reveals distinct patterns across the classifiers. ResNet18 and MobileNetV2 demonstrate similar resilience, with success rates converging around 74\% and 73\% respectively. In contrast, DenseNet121 shows the highest vulnerability, with rates reaching approximately 83\% across all scenarios, challenging the notion that more complex models offer better protection against adversarial attacks. While scenarios eventually converge to similar lifetime success rates for all classifiers, their learning trajectories differ. Initially, scenarios providing more information (True Distribution and Correct Confidence Only) lead to higher success rates across all models, highlighting the impact of information disclosure on attack effectiveness.

\subsection{Efficiency of Attacks Across Classifiers}

\begin{figure}[t]
\centering
\begin{subfigure}[b]{0.45\textwidth}
    \caption{ResNet18}
    \label{fig:resnet18_steps}
    \includegraphics[width=\textwidth]{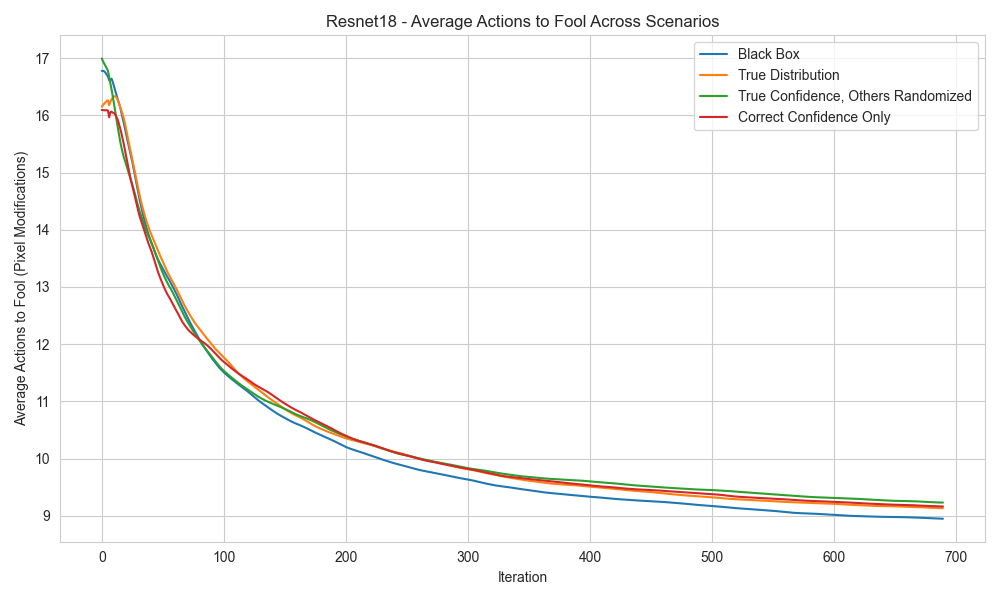}
\end{subfigure}

\begin{subfigure}[b]{0.45\textwidth}
    \caption{DenseNet121}
    \label{fig:densenet121_steps}
    \includegraphics[width=\textwidth]{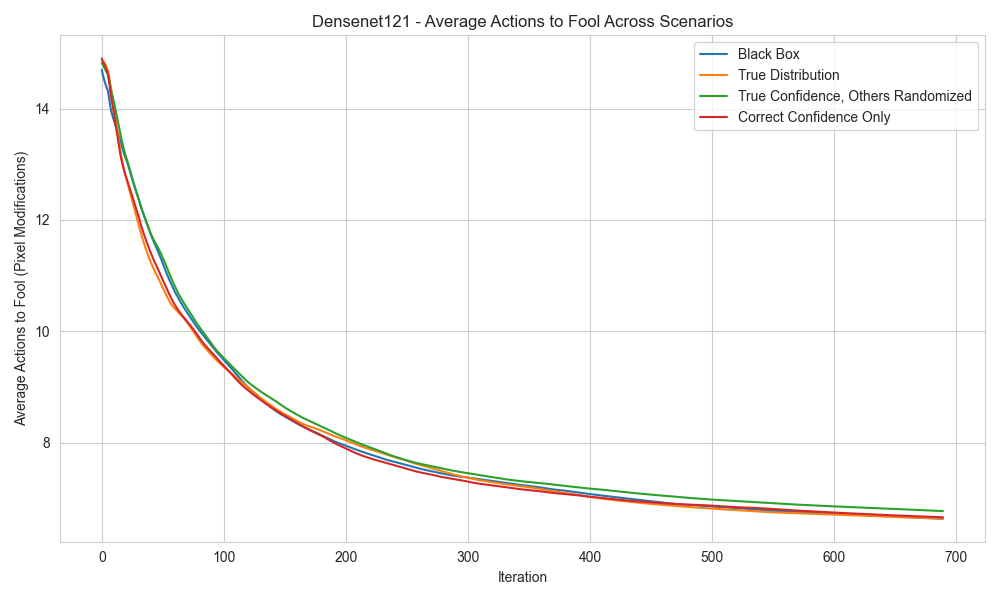}
\end{subfigure}

\begin{subfigure}[b]{0.45\textwidth}
    \caption{MobileNetV2}
    \label{fig:mobilenetv2_steps}
    \includegraphics[width=\textwidth]{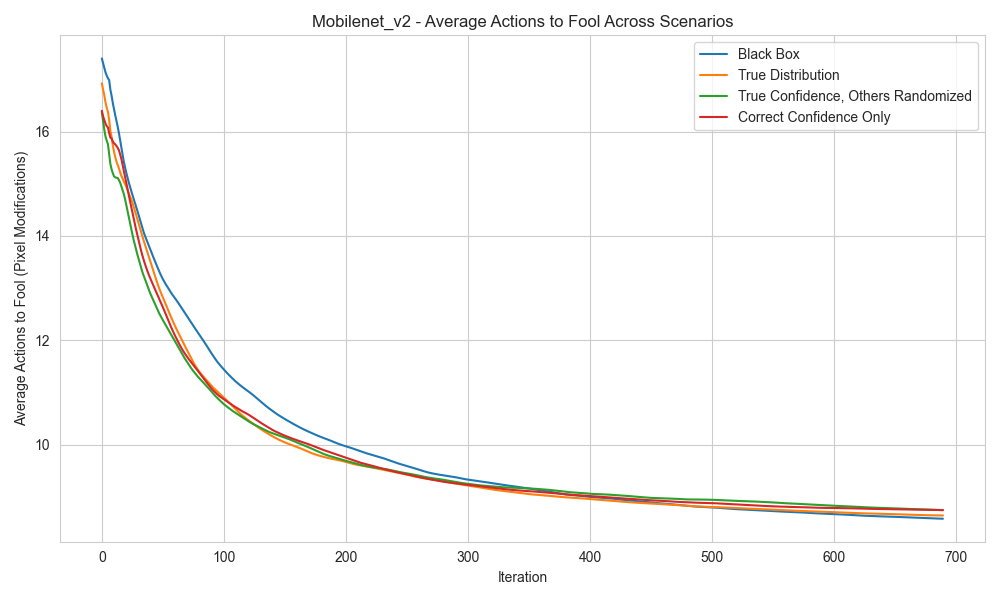}
\end{subfigure}

\caption{Averaged AAF over the 43 classes for different attack scenarios across classifiers.}
\label{fig:steps_to_fool}
\end{figure}

The average AAF required for successful attacks varies significantly across classifiers and scenarios, as shown in Fig. \ref{fig:steps_to_fool}.

DenseNet121 requires the fewest modifications for successful attacks, averaging about 6.5 pixel changes. MobileNetV2 needs an average of 8.6 pixel changes, while ResNet18 proves to be the most resilient, needing about 9.1 pixel changes on average for a successful attack.

\subsection{Impact of Different Scenarios Across Classifiers}

\begin{figure}[t]
\centering
\begin{subfigure}[b]{0.43\textwidth}
    \caption{ResNet18}
    \includegraphics[width=\textwidth]{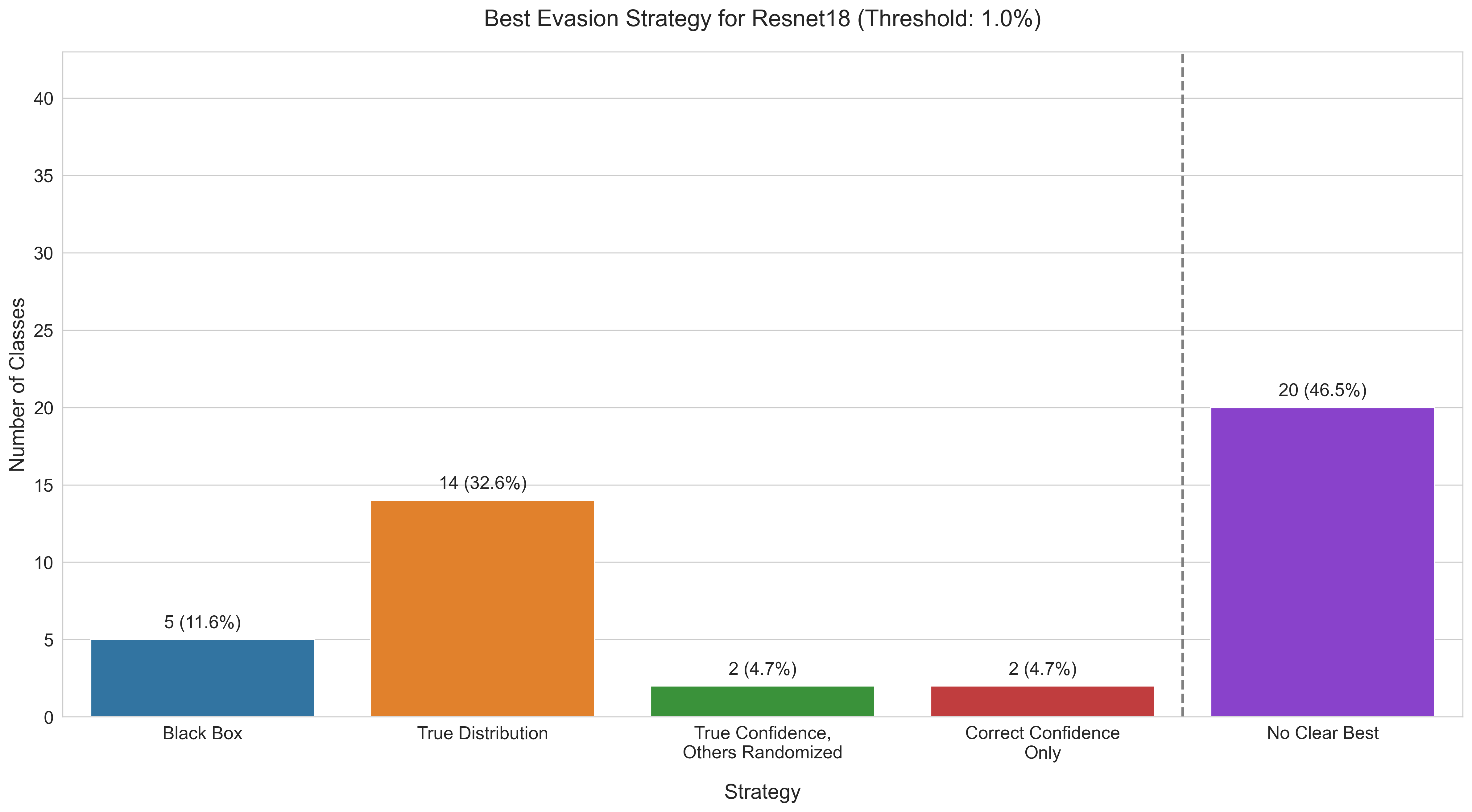}
    \label{fig:resnet18_counts}
\end{subfigure}
\begin{subfigure}[b]{0.43\textwidth}
    \caption{DenseNet121}
    \includegraphics[width=\textwidth]{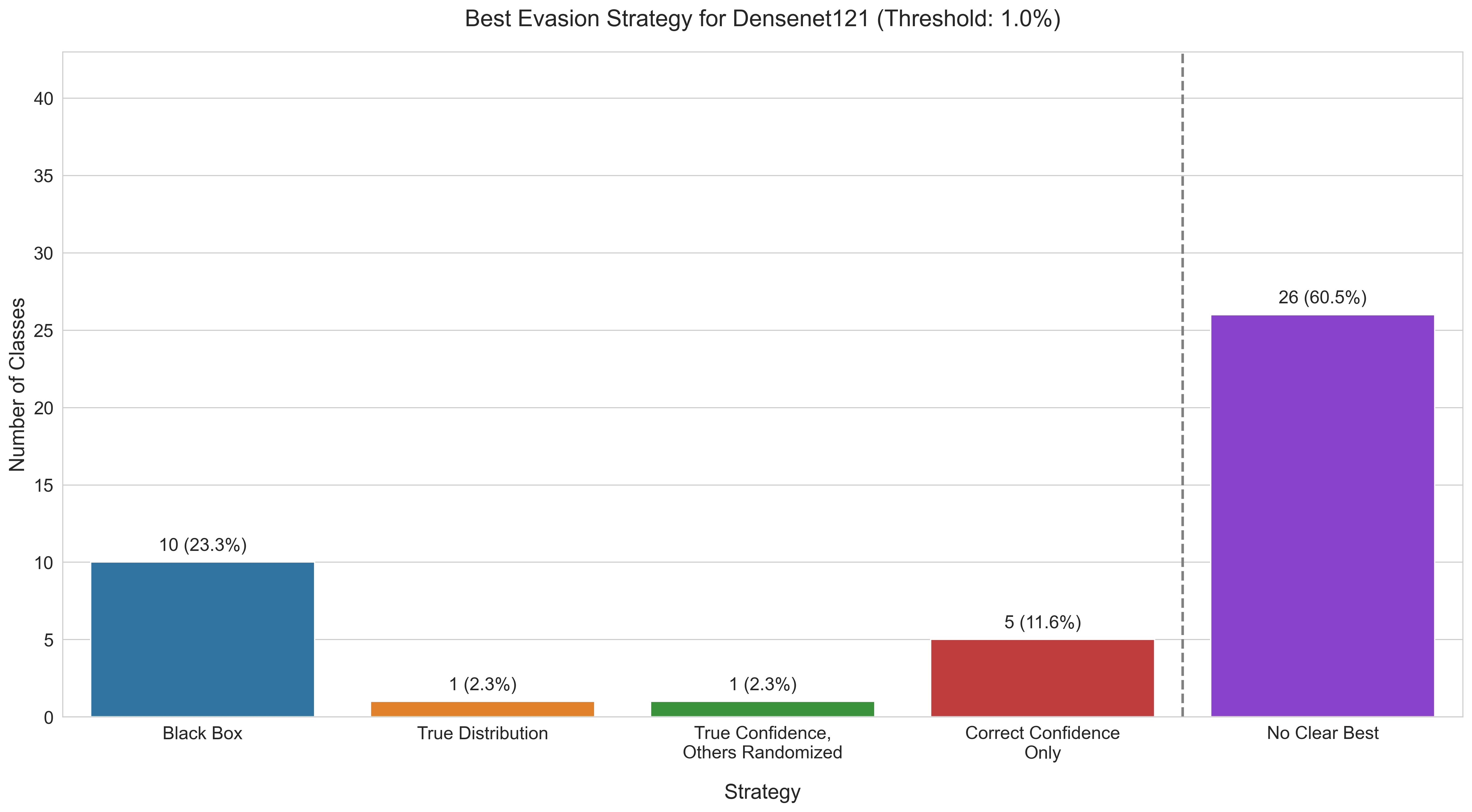}
    \label{fig:densenet121_counts}
\end{subfigure}
\begin{subfigure}[b]{0.43\textwidth}
    \caption{MobileNetV2}
    \includegraphics[width=\textwidth]{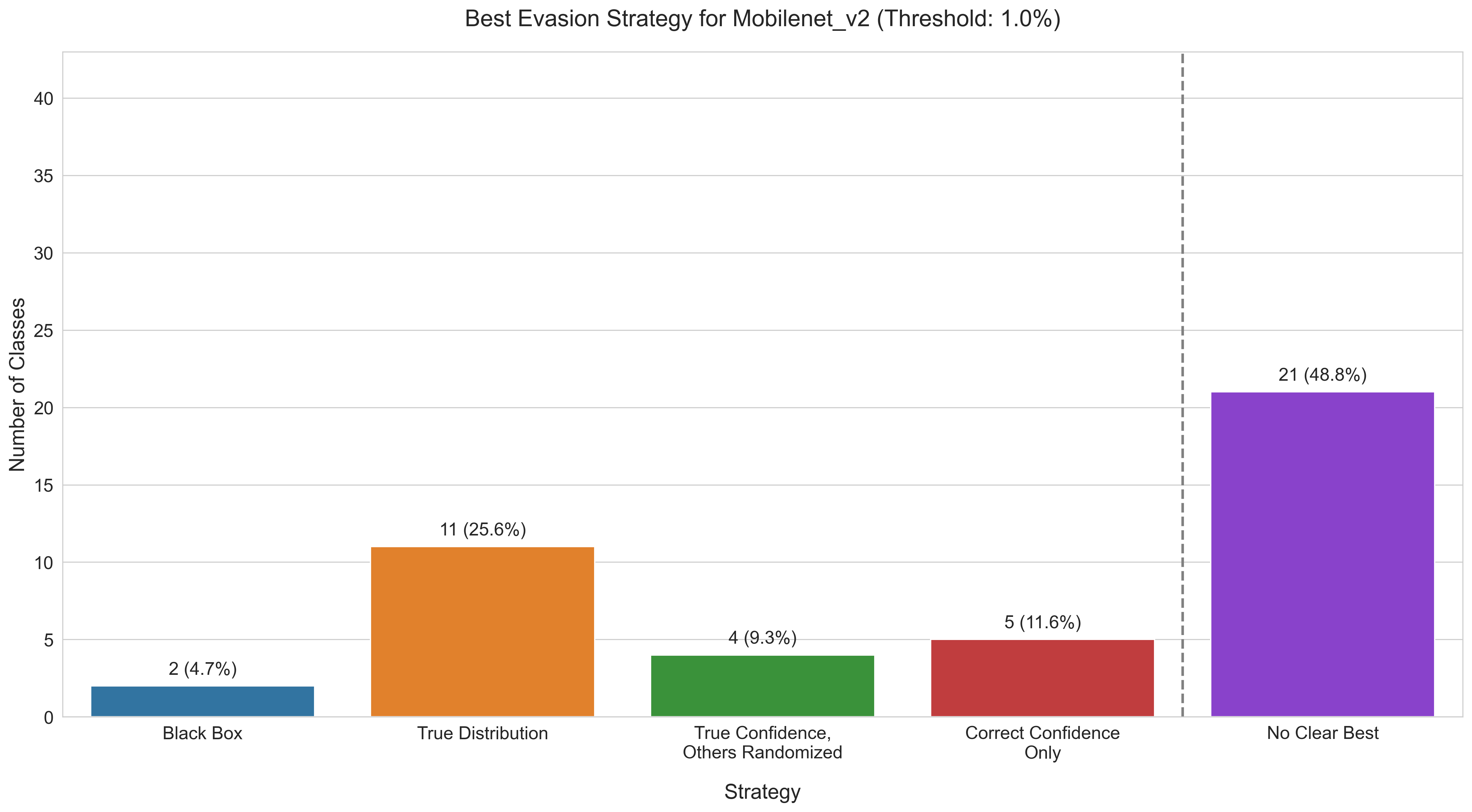}
    \label{fig:mobilenetv2_counts}
\end{subfigure}

\caption{Distribution of most effective attack scenarios across classes for each classifier. Each bar represents the count of classes (out of 43 total) where a specific scenario resulted in the highest LSR for the attacker, exceeding other scenarios by at least 1\%. Percentages are calculated out of 43 total classes. The No Clear Best category indicates cases where multiple strategies perform similarly within the threshold.}
\label{fig:best_strategy_counts}
\end{figure}

Fig. \ref{fig:best_strategy_counts} illustrates the distribution of most effective attack scenarios across classes for each classifier.

To construct these charts, we analyzed the performance of each attack scenario on every individual class using the LSR metric. We introduced a threshold parameter $t = 0.01$ (1\%) to focus on meaningful differences. A scenario is considered the "best" for a given class only if its LSR exceeds that of all other scenarios by at least $t$. The bars in the charts represent the count of classes (out of 43 total) for which each scenario had the highest LSR under this criterion. It's important to note that No Clear Best is not an attack scenario, but rather a category indicating that all scenarios had LSRs within the $t$ threshold of each other for those classes. 

\textbf{From the attacker's perspective:} The goal is to identify which scenarios are most effective across classes. In this context:

\begin{itemize}
    \item The No Clear Best category (purple bar) represents classes where all strategies achieve similar LSRs (within the $t$ threshold).
    \item The other bars indicate scenarios that outperform others in more classes. Higher bars are preferable for the attacker, as they represent more classes where that strategy is notably more effective.
    \item An ideal outcome for the attacker would show high bars for specific scenarios, indicating clear superiority in many classes.
\end{itemize}

\textbf{From the defender's perspective:} The goal is to identify which scenarios limit the attacker's effectiveness across classes. In this context:

\begin{itemize}
    \item The No Clear Best category represents classes where no defensive strategy offers a significant advantage over others in terms of limiting the attacker's LSR.
    \item Lower bars for specific scenarios are preferable, as they represent fewer classes where that scenario is notably more effective for the attacker.
    \item An ideal outcome for the defender would show low bars across all specific scenarios, indicating no single scenario is consistently advantageous for the attacker.
\end{itemize}

\begin{figure}[b]
\centering
\includegraphics[width=\linewidth]{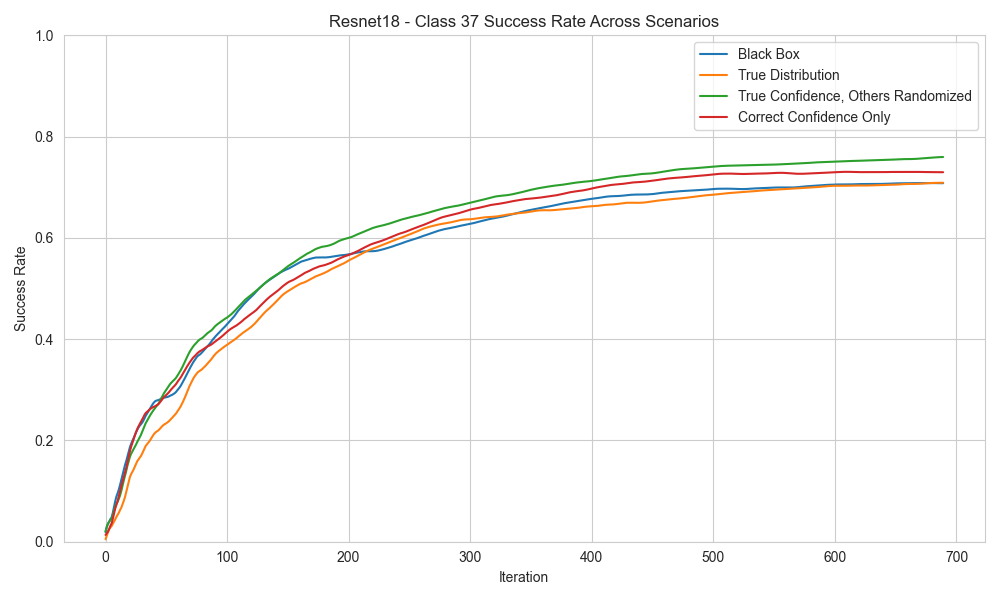}
\caption{ResNet18: LSRs across different attack scenarios for Class 37.} \label{fig:resnet18_class37_success}
\end{figure}

For ResNet18, the True Distribution scenario was most advantageous for the attacker in 14 classes, while the Black Box scenario was best in 5 cases. The True Confidence, Others Randomized and Correct Confidence Only scenarios were each most effective for the attacker in 2 classes. 20 out of 43 classes showed no clear best strategy for the attacker. Fig. \ref{fig:resnet18_class37_success} illustrates the success rates across different attack scenarios for Class 37 in ResNet18. For this class, the True Confidence, Others Randomized scenario shows a success rate approximately 5\% higher than the other scenarios. For the other class where True Confidence, Others Randomized was most effective, it showed approximately a 1.5\% higher LSR compared to other scenarios.

For DenseNet121, the Black Box scenario was most advantageous for the attacker in 10 out of 43 classes. The True Distribution and True Confidence, Others Randomized scenarios were each optimal for the attacker in 1 class. In the class where True Confidence, Others Randomized was optimal, it showed an approximately 1.5\% higher LSR compared to other scenarios.

For MobileNetV2, the True Distribution scenario was most beneficial for the attacker in 11 classes. The noise-based defense (True Confidence, Others Randomized) was the best scenario in four classes (35, 38, 39, and 41). In class 39, this scenario increased the attacker's success rates by up to 30\% compared to the Black Box scenario, and by approximately 16-17\% compared to the other scenarios, as shown in Fig. \ref{fig:mobilenetv2_class39}. The Black Box scenario was optimal for the attacker in 2 classes.

\subsection{Adversarial Training Loop in MobileNetV2}

\begin{figure}[b]
\centering
\begin{minipage}{0.20\linewidth}
    \centering
    \includegraphics[width=\linewidth]{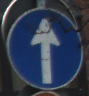}
    \caption*{Class 35}
\end{minipage}
\hfill
\begin{minipage}{0.20\linewidth}
    \centering
    \includegraphics[width=\linewidth]{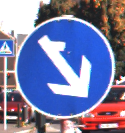}
    \caption*{Class 38}
\end{minipage}
\hfill
\begin{minipage}{0.20\linewidth}
    \centering
    \includegraphics[width=\linewidth]{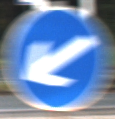}
    \caption*{Class 39}
\end{minipage}
\hfill
\begin{minipage}{0.20\linewidth}
    \centering
    \includegraphics[width=\linewidth]{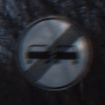}
    \caption*{Class 41}
\end{minipage}
\caption{Sample images from classes where using noise as a defense in MobileNetV2 benefits the attacker.}
\label{fig:noise_sample_images}
\end{figure}

\begin{figure}[t]
\centering
\includegraphics[width=\linewidth]{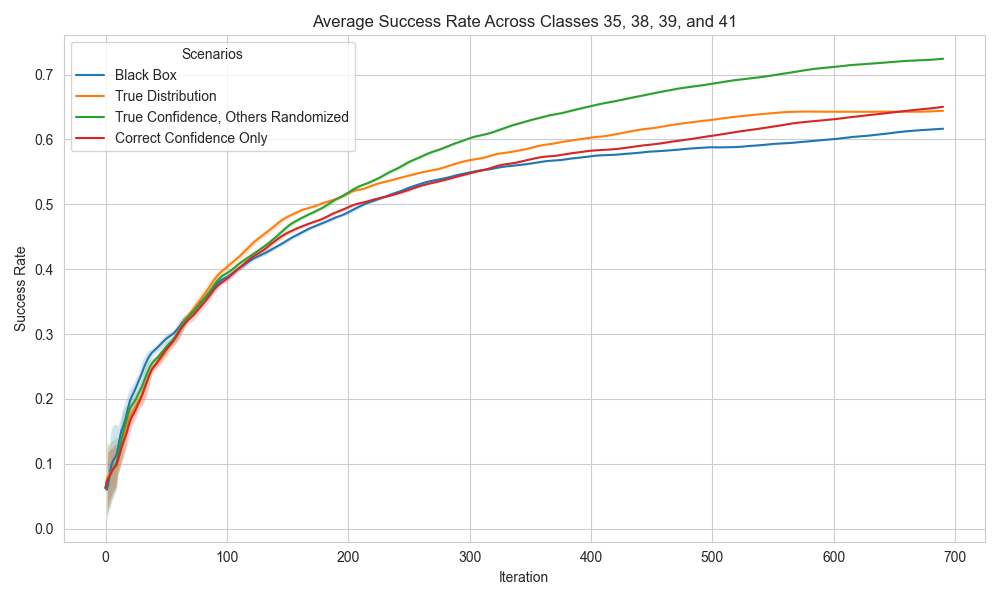}
\caption{MobileNetV2: LSRs for classes 35, 38, 39, and 41 where the noise-based defense backfires.}
\label{fig:mobilenetv2_noise_samples}
\end{figure}

A unique and significant phenomenon was observed with MobileNetV2, where the noise-based defense scenario appeared to create an adversarial training loop. This effect was particularly pronounced in classes 35, 38, 39, and 41, which are characterized by visually noisy images, as shown in Figs. \ref{fig:noise_sample_images} and \ref{fig:mobilenetv2_noise_samples}.

\begin{figure}[b]
\centering
\includegraphics[width=\linewidth]{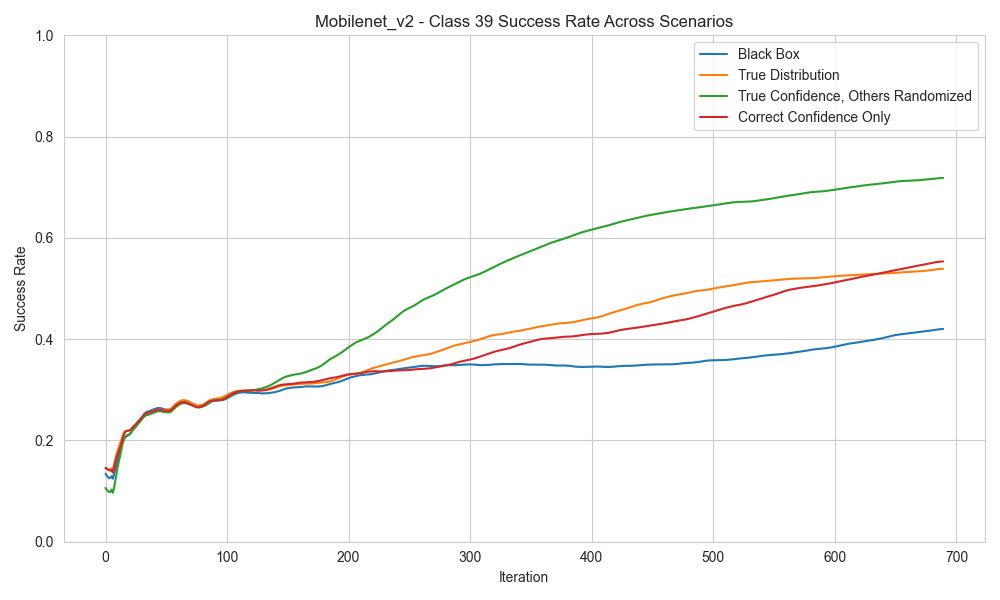}
\caption{MobileNetV2: LSRs of the attack scenarios on Class 39, demonstrating the significant backfire effect of the noise-based defense.}
\label{fig:mobilenetv2_class39}
\end{figure}

For these classes in MobileNetV2, the True Confidence, Others Randomized scenario consistently outperformed other strategies, including those providing more accurate information. However, the effect was most striking for Class 39, as illustrated in Fig. \ref{fig:mobilenetv2_class39}.

For Class 39 in MobileNetV2, the noise-based defense led to a dramatic increase in the agent's evasion success rate. The True Confidence, Others Randomized scenario outperformed other strategies by approximately 16-17\%. Moreover, the gap between this noise-based scenario and the Black Box scenario reached up to 30\%, highlighting the significant vulnerability introduced by the noise-based defense for this particular class.

This extreme case in Class 39 highlights the potential risks of employing noise-based defenses without considering their impact on specific classes or image types. It suggests that for certain visual patterns or decision boundaries in MobileNetV2, the introduction of noise in confidence values can create exploitable patterns that significantly benefit the attacker.

This phenomenon was not observed to the same extent in DenseNet121 or ResNet18, suggesting that the creation of this adversarial training loop may be dependent on specific architectural features of MobileNetV2. The lightweight nature of MobileNetV2, designed for efficiency, may inadvertently make it more susceptible to this type of exploitation when noise is introduced into the confidence values.

\section{Discussion}

Our results reveal a complex, often counterintuitive relationship between information disclosure scenarios, noise-based defenses, and RL-based evasion attacks across various classifier architectures. These findings challenge common assumptions in adversarial machine learning and highlight the need for nuanced, model-specific defense strategies.

One striking result is the effectiveness of the Black Box scenario in DenseNet121, outperforming other strategies in 23.3\% of classes. This challenges the assumption that more information invariably aids the attacker, suggesting that in some architectures, additional information might not enhance attack performance. While it's premature to draw definitive conclusions, these results indicate that the impact of information disclosure on attack success may be more complex and model-dependent than previously thought.

The varying impact of the noise-based defense (True Confidence, Others Randomized scenario) across models and classes is another key finding. As shown in Fig. \ref{fig:best_strategy_counts}, this scenario backfired in four classes for MobileNetV2, two for ResNet18, and one for DenseNet121. This variability hightlights the importance of considering model architecture when implementing noise-based defenses. The backfire effect could be attributed to the RL agent's ability to exploit the introduced randomness as an additional information source, particularly in classes with less distinct decision boundaries.

Figures \ref{fig:resnet18_lifetime}, \ref{fig:densenet121_lifetime}, and \ref{fig:mobilenetv2_lifetime} demonstrate that noise-based defenses can initially buy time against adversarial attacks. However, a crucial observation is that across all three classifiers, the LSRs for different scenarios converge to within 1-2\% of each other by the end of the training period. This indicates that while noise-based defenses can delay the attacker's progress, they don't provide long-term protection against a persistent, adaptive RL-based attacker.

The backfire effect is particularly evident in MobileNetV2 for class 39, where the noise-based defense becomes the most advantageous scenario for the attacker, with up to a 30\% higher success rate compared to the Black Box scenario. Similarly, for ResNet18 in class 37 (Fig. \ref{fig:resnet18_class37_success}), the noise-based defense becomes the most beneficial scenario, with a 5\% higher success rate.

These results demonstrate the RL agent's remarkable capability to adapt and overcome noise-based defenses, eventually achieving success rates comparable to or exceeding scenarios with more complete information. This points to the need for more robust and dynamic defensive strategies in adversarial machine learning.

The creation of an apparent adversarial training loop, particularly in MobileNetV2 for visually noisy classes, raises important questions about the relationship between model architecture, size, and vulnerability to this backfire effect. The effect was most pronounced in MobileNetV2 (the smallest model), less so in DenseNet121 (middle-sized), and limited in ResNet18 (largest). This pattern suggests that model size, architecture, and the nature of the data (particularly in noisy classes) interact to create conditions where noise-based defenses can backfire.

As the industry moves towards edge computing and smaller models, these findings raise concerns about potential security implications. While our results show a correlation between model size and susceptibility to this effect, it remains an open question whether this is directly caused by model compression or if other architectural features play a more significant role.

The RL agent's high success rates across all models, even in the Black Box scenario, demonstrate the power of reinforcement learning in developing adaptive attack strategies. This suggests that traditional defenses, including those based on randomness or information hiding, may need re-evaluation in the context of sophisticated, learning-based attacks.

These results emphasize the need for a more nuanced, model-specific approach to defensive strategies in AML. While noise-based defenses can be effective against static attacks and can buy time against adaptive attackers, they may introduce long-term vulnerabilities, particularly in certain architectures like MobileNetV2. Future defense mechanisms should consider not only the potential for attackers to exploit introduced randomness or provided information but also how this exploitation may vary across different model architectures, sizes, and classes.

As we progress towards more efficient, compressed models for edge computing, understanding and mitigating the potential security implications of these architectural choices becomes crucial. Our findings suggest that the relationship between model compression and adversarial vulnerability is complex and warrants further investigation. Developing new defensive strategies specifically tailored to smaller, more efficient models may be necessary to ensure their robustness against adaptive adversarial attacks.

\section{Conclusion and Future Work}

This work demonstrates a counterintuitive phenomenon in adversarial machine learning: the potential for noise-based defenses to inadvertently benefit RL-based attackers, particularly when dealing with noisy images. Our study across multiple classifier architectures reveals that the effectiveness of defensive strategies varies significantly, challenging the notion of a one-size-fits-all approach to adversarial defense.

ResNet18 demonstrated best performance when providing only the correct confidence or using a noise-based defense, again showing vulnerability when full distribution information was disclosed. For DenseNet121, we found that providing either the true confidence distribution or employing a noise-based defense proved most effective, while limiting information through a Black Box scenario was least effective. Conversely, MobileNetV2 showed highest resilience under the Black Box scenario, with full information disclosure being the most vulnerable state. 

These findings challenge the conventional wisdom that randomness or information limitation is universally beneficial as a defense against adversarial attacks. We've shown that in certain cases, especially with visually noisy data, the effectiveness of these strategies can vary dramatically based on the underlying model architecture. This showcases the complexity of securing machine learning systems and the need for nuanced, context-aware defensive strategies.

The discovery of potential adversarial training loops, particularly evident in MobileNetV2 with visually noisy classes, highlights a critical area for future research. As machine learning systems become increasingly central to critical applications, it's vital to develop defenses that can adapt to and counteract reinforcement learning-based attackers capable of learning from and exploiting various levels of information disclosure.

Future work should focus on understanding the long-term dynamics of these adversarial interactions across a broader range of model architectures and datasets. Investigating why certain architectures are more susceptible to specific defensive strategies could provide valuable insights for designing more robust models. Additionally, exploring alternative randomization techniques and their impact on different model architectures could lead to more effective defensive strategies.

This study opens several avenues for future research, including the evaluation of our methodology against datasets with varying levels of inherent noise, and the incorporation of confidence values into the reward function to potentially enhance the agent's evasion capabilities. While this study used the Dirichlet distribution for noise-based defense, investigating other distributions such as truncated Gaussian or Beta could provide insights into optimal noise-based defense strategies for different scenarios and model architectures.

In conclusion, our work highlights the complex link between model architecture, information disclosure, and adaptive attack strategies in adversarial machine learning. It emphasizes the need for defensive approaches that are carefully tailored to specific model architectures and capable of adapting to the evolving strategies of RL-based attackers. As the field of adversarial machine learning continues to evolve, studies like this will be crucial in developing the next generation of secure and robust AI systems across various model architectures and application domains.

\bibliographystyle{IEEEtran}
\bibliography{references}

\end{document}